# A two-fluid helicity transport model for flux-rope merging

**S You**

Aeronautics & Astronautics, University of Washington, Seattle, WA 98195, USA

Email: syou@aa.washington.edu

**Abstract**

A two-fluid model for canonical helicity transport between merging flux-ropes is developed. Relative canonical helicity is defined here as the gauge invariant helicity of the canonical momentum of a given species, and represents the generalization of helicity to magnetized plasmas with non-negligible momentum. The model provides the conditions for when the transport of canonical helicity during merging manifests itself as magnetic helicity transfer or as generation of strong helical plasma flows. The model is applied to compact torus merging but is also relevant to situations where magnetic flux-ropes and flow vorticity flux-ropes interact. Geometrical interpretations are given.

## 1. Introduction

Magnetic flux-rope dynamics in plasmas can be interpreted geometrically with images of twists, writhes and links transferred across individual reconnecting flux tubes [1, 2, 3, 4]. Global behaviour is expected to be constrained by magnetic helicity conservation [5, 6], provided the assumptions behind ideal magnetohydrodynamics (MHD) holds everywhere except within the small reconnection regions. Because these assumptions freeze the plasma to the magnetic flux, from a topological point-of-view, there is no need to distinguish between the behaviour of the plasma fluid and the shape of magnetic fields. However, non-ideal physics within a small region clearly affects the global idealized behaviour far from these regions [7, 8, 9]. For example, laboratory experiments observe that non-ideal physics in a reconnection volume affects the global topology of magnetized low-beta plasmas [10], spontaneous global plasma flows develop from non-linear saturation of magnetic turbulence [11] or internal reconnection events [8, 12], and localized shear flows can stabilize global plasma instabilities [13]. Any local electromagnetic or particle behaviour originating from non-ideal physics must, and will, propagate through the boundaries of the small volume under consideration.

This paper demonstrates that a helicity transport model built from two-fluid magnetohydrodynamics governs the dynamics of magnetized plasma fluids. The model shows that the topological property of interest, canonical helicity, can be transferred from one fluid to another fluid while preserving the total canonical helicity. Because canonical flux tubes are the weighted sum of vorticity flux tubes and magnetic flux tubes, helicity transport effectively results in changes in plasma flows or magnetic fields. The model provides the conditions for helicity tranfer, and in particular, when helicity injection would result in plasma flows or in helical magnetic fields. The model provides a more fundamental constraint for the dynamics of plasmas than simple magnetic helicity conservation, which is only relevant to quasi-static systems. The model retrieves all previous results for static ideal MHD plasmas [3, 5, 6, 14] and ordinary neutral fluids [15] in the appropriate limits, while extending the same topological concepts to regimes where flowing two-fluid plasmas are applicable [16, 17, 18]. The model provides the conditions for when the transport of canonical helicity during merging of flux ropes manifests itself as magnetic helicity transfer or as strong helical plasma flows. The model presented here does not consider relaxation arguments which can determine the final flow and magnetic profiles [19, 20, 18, 17, 7] nor on turbulent mean-field effects calculated in a self-consistent manner with energy evolution [12], but generalizes the intuitive geometric interpretation of helicity transport [1, 2, 21] to include flows.

Section 2 begins with definitions of relative canonical helicity. Previous generalizations of helicity [20, 18, 7, 22] have ignored situations where canonical flux tubes can intercept the boundaries of a given volume and, even within an isolated system, ignored the fact that electron canonical flux tubes necessarily intercept ion canonical flux tubes. These earlier attempts at generalizations of helicity have only considered ordinary helicity without accounting for gauge invariance. Section 3 presents a geometrical interpretation of canonical helicity which helps visualize the evolution (section 4) due to enthalpy changes on boundary surfaces, time-dependent forces or various sinks. Section 5 provides the two-fluid model for flux rope merging based on these equations, applying it to a flux-rope merging experiment and section 6 interprets the same model from an equivalent MHD point-of-view before concluding in section 7.

## 2. Definitions

Canonical helicity $K_\sigma$ is calculated from the canonical momentum $\vec{P}_\sigma = m_\sigma \vec{u}_\sigma + q_\sigma \vec{A}$ of a fluid element of species $\sigma$, with mass $m_\sigma$ and charge $q_\sigma$, flowing at velocity $\vec{u}_\sigma$ in a magnetic field $\vec{B} = \nabla \times \vec{A}$ determined by the vector potential $\vec{A}$, and its circulation or canonical vorticity $\vec{\Omega}_\sigma = \nabla \times \vec{P}_\sigma$, summed over the complete volume $V_\sigma$ under consideration as

$$K_{\sigma rel} = \int_{V_\sigma} \vec{P}_{\sigma-} \cdot \vec{\Omega}_{\sigma+} \, dV \qquad (1)$$

where the subscripts $rel$ indicates (gauge invariant) relative canonical helicity, and the plus and minus subscripts are shorthands [3] for a given scalar $x$ or vector field $\vec{X}$ offset by a reference field $x_{ref}$ or $\vec{X}_{ref}$, i.e. $x_\pm = x \pm x_{ref}$ or $\vec{X}_\pm = \vec{X} \pm \vec{X}_{ref}$. Earlier versions of equation (1) variously called this quantity generalized vorticity [7], self-helicity [18], generalized helicity [20, 12], or fluid helicity [23], which can be confused with ordinary fluid helicity [15], itself often known as kinetic helicity. To alleviate confusion, this paper uses the term canonical helicity [24] when considering the helicity of the vector field $\vec{\Omega}_\sigma$. To eliminate gauge dependence, the arbitrary hypothetical reference fields must satisfy the conditions $\vec{X}_{ref} \cdot d\vec{S} = \vec{X} \cdot d\vec{S}$ or $\nabla x_{ref} \cdot d\vec{S} = \nabla x \cdot d\vec{S}$ at each unit surface with normal $d\vec{S}$ on the surface $S_\sigma$ bounding the volume $V_\sigma$, and with circulation equal to the actual fields in the volume external to the volume of interest, i.e $\vec{X}_{ref} = \vec{X} + \nabla x$. The subscript $\sigma$ appends the symbols for the volume and surface under consideration to emphasize that the integration is performed separately for each fluid and that relative helicity must be considered because the fluid volume of one species intersects the fluid volume of the other species at some common surfaces.

A convenient point-of-view expands equation (1) to the weighted sum of three helicities

$$K_{\sigma rel} = m_\sigma^2 \mathcal{H}_{\sigma rel} + m_\sigma q_\sigma \mathcal{X}_{\sigma rel} + q_\sigma^2 \mathcal{K}_{rel} \qquad (2)$$

where $\mathcal{H}_{\sigma rel} = \int \vec{u}_{\sigma-} \cdot \vec{\omega}_{\sigma+} \, dV$ is the fluid relative kinetic helicity (fluid vorticity is $\vec{\omega}_\sigma = \nabla \times \vec{u}_\sigma$), $\mathcal{X}_{\sigma rel} = \int (\vec{u}_{\sigma-} \cdot \vec{B}_+ + \vec{u}_{\sigma+} \cdot \vec{B}_-) \, dV$ is the relative cross-helicity and $\mathcal{K}_{rel} = \int \vec{A}_- \cdot \vec{B}_+ \, dV$ is the relative magnetic helicity. If the mass of the electron can be ignored, then electron canonical helicity is simply magnetic helicity weighted by the electrical charge $e^2$ and equation (2) for the ion (plasma) fluid becomes $K_i = m_i^2 \mathcal{H}_i + m_i q_i \mathcal{X}_i + q_i^2 \mathcal{K}$. The total global canonical helicity can then be defined as $\mathbb{K}_{rel} = \sum_\sigma K_{\sigma rel}$ and fundamentally represents the comprehensive helicity of multi-species magnetized plasmas.

## 3. Geometrical interpretation of canonical helicity

Consider a collimated cylindrical current-carrying magnetic flux tube. The vector fields on an arbitrary flux surface of this tube are drawn in Figure 1. There are three possible free parameters for the system for which we choose the following. The first is the current density which we take to be a fixed axial current $J_z$ only, since the system is fixed with no time-changing flux, we can suppose that $J_\theta$ is zero. The second is the axial magnetic field $B_z$ and because the tube is collimated, the axial magnetic $B_z$ field can be taken to be a constant. The third is the electron fluid flow velocity $\vec{u}_e$. If we consider the electrons as massless, then they are tied to the magnetic field lines but their velocity will be determined by some parallel electric field or pressure gradients in the equation of motion $\vec{u}_e = u_e \hat{b}$ where $\hat{b} = \vec{B}/|\vec{B}|$. Because the current density is $\vec{J} = ne(\vec{u}_i - \vec{u}_e)$ then the ion fluid flow velocity is $\vec{u}_i = \vec{u}_e + \vec{J}/ne$. Since the system is supposed to be azimuthally symmetric and axially symmetric, the magnetic vector potential becomes $\{A_\theta = rB_z/2; A_z = -rB_\theta/2\}$ and the canonical vorticity becomes $\{\Omega_z = P_\theta/2; \Omega_\theta = -P_z/r\}$. Figure 1(b) shows an example vector field with the three free parameters set to $J_z = 8$, $B_z = 7$, and $u_e = 4$ and all physical constants set to 1. The total magnetic field is helical because of the superposition of axial currents with an axial magnetic field. The electrons are massless so they flow along the magnetic field, and therefore in a helical manner. The flow is assumed here to be parallel and may be due for example to a steady electrostatic potential, neglecting any pressure gradients or time dependent magnetic vector potential. The ion flow must therefore compensate for the electron direction to make up the axial current density. Because the current density is purely axial, the ions and electrons must flow together azimuthally at the same velocity (no net azimuthal current). The ion flow is therefore helical. The canonical momenta are also helical as are the canonical vorticities and supposing the system is symmetric, the total helicity need not be zero. The collimated current-carrying magnetic flux tube is thus identically described as a twisted flux tube of electron canonical vorticity (effectively an anti-parallel magnetic flux tube) intertwined with a twisted flux tube of ion canonical vorticity as represented in Figure 1(a). Because of this intertwining, any change to the helicity of one tube can affect the helicity of the other tube, given the appropriate conditions for the transport of links, writhes, twists and crossings. This topological argument is manifested physically as changes in magnetic field line pitch and helical plasma flows. With the appropriate inductive or enthalpy boundary conditions, magnetic field line pitch can decrease to increase helical plasma flows or, conversely, flow line pitch could decrease to increase the helicity of magnetic field lines.

## 4. Transport of canonical helicity

The evolution of canonical helicity is based on the two-fluid equations of motion expressed in canonical form [24, 2, 18] as

$$\frac{\partial \vec{P}_\sigma}{\partial t} + \vec{\Omega}_\sigma \times \vec{u}_\sigma = -\nabla h_\sigma - \frac{\vec{R}_\sigma}{n_\sigma} \quad (3)$$

where $h_\sigma \equiv q_\sigma \phi + 1/2\, m_\sigma u_\sigma^2 + \int_0^{\mathcal{P}_\sigma} d\mathcal{P}_\sigma/n_\sigma$ is the enthalpy of the plasma that combines the work done on the plasma in conservative fields, ie. from an electrostatic potential $\phi$, mechanical work and scalar pressure $\mathcal{P}_\sigma$. Gravitational and other terms can be included if necessary [18]. The general non-conservative friction force $\vec{R}_\sigma = \vec{R}_{\sigma\alpha} + \vec{R}_{\sigma\sigma} + \vec{R}_{th\sigma}$ combines the interspecies drag $\vec{R}_{\sigma\alpha} = \nu_{\sigma\alpha} n_\sigma m_\sigma (\vec{u}_\sigma - \vec{u}_\alpha)$ due to collisions between species $\sigma$ and $\alpha$ at frequency $\nu_{\sigma\alpha}$, viscous drag components $\vec{R}_{\sigma\sigma} = \nabla \cdot (\overline{\overline{\Pi}}_\sigma - \mathcal{P}_\sigma \hat{I})$ of the pressure tensor $\overline{\overline{\Pi}}_\sigma$ and any thermal Nernst effect $\vec{R}_{th\sigma}$. The effect of energy on the system occurs on the conservative enthalpy term $\nabla h_\sigma$ and the non-conservative friction term $\vec{R}_\sigma$ which are kept deliberately general. Self-consistent closure requires an energy balance equation of state [12, 25], but for the purposes of helicity evolution, the effects of energy are only "seen" as changing enthalpies or friction. The analysis here focuses on the evolution of canonical helicity from given boundary conditions rather than on the self-consistent effects of helicity evolution on energy evolution. Defining a canonical electric field $\vec{\mathbb{E}}_\sigma \equiv -\nabla h_\sigma - \partial \vec{P}_\sigma/\partial t$ by direct analogy with the real electric field $\vec{E} = -\nabla \phi - \partial \vec{A}/\partial t$ simplifies equation (3) to a canonical Ohm's law

$$\vec{\mathbb{E}}_\sigma + \vec{u}_\sigma \times \vec{\Omega}_\sigma = \frac{\vec{R}_\sigma}{n_\sigma} \quad (4)$$

of which the curl becomes an induction equation or Faraday's law for canonical quantities $\nabla \times \vec{\mathbb{E}}_\sigma = -\partial \vec{\Omega}_\sigma/\partial t$. Equation (4) shows that in the absence of friction, canonical vorticity is frozen to the species and if friction is dominated only by interspecies collisions $\vec{R}_\sigma \sim \vec{R}_{\sigma\alpha} = -\vec{R}_{\alpha\sigma}$ then canonical vorticity can be transferred across species.

For gauge invariant relative helicity, the reference version of equation (3) must be used,

$$\frac{\partial \vec{P}_{\sigma-}}{\partial t} - (\vec{u}_\sigma \times \vec{\Omega}_\sigma - \vec{u}_{\sigma ref} \times \vec{\Omega}_{\sigma ref}) = -\nabla h_{\sigma-} - \frac{\vec{R}_{\sigma-}}{n_\sigma} \quad (5)$$

with the reference enthalpy chosen to be $h_{\sigma ref} \equiv q_\sigma \phi_{ref} + 1/2\, m_\sigma u_{\sigma ref}^2 + \int_0^{\mathcal{P}_{\sigma ref}} d\mathcal{P}_\sigma / n_\sigma$ in order to have $h_{\sigma-}$ be consistent with the thermodynamic definition of enthalpy (a relative quantity by definition) [24]. From this choice of enthalpy, the reference canonical momentum field is $\vec{P}_{\sigma ref} = m_\sigma \vec{u}_{\sigma ref} + q_\sigma \vec{A}_{ref}$ which is constructed from a chosen reference flow field $\vec{u}_{\sigma ref}$ and reference magnetic field $\vec{A}_{ref}$. The choice of reference for relative canonical helicity is thus fully specified from the four fields $\phi_{ref}, \vec{A}_{ref}$ and $\vec{u}_{\sigma ref}$.

The evolution of canonical helicity is derived using the definition of the canonical electric field and and canonical Faraday's law to give

$$\frac{dK_{\sigma rel}}{dt} = -\int_V (\vec{\mathbb{E}}_{\sigma+} \cdot \vec{\Omega}_{\sigma-} + \vec{\mathbb{E}}_{\sigma-} \cdot \vec{\Omega}_{\sigma+}) dV - \int_S h_{\sigma-} \vec{\Omega}_{\sigma+} \cdot d\vec{S} - \int_S \vec{P}_{\sigma-} \times \frac{\partial \vec{P}_{\sigma+}}{\partial t} \cdot d\vec{S} \qquad (6)$$

after using the solenoidal property of the canonical vorticity, and ignoring the Leibniz convection term associated with the motion of boundaries. The first term on the right-hand side is the generalization of the $\int \vec{E} \cdot \vec{B}\, dV$ term in the evolution of magnetic helicity and is easily simplified to $2 \int (\vec{R}_{\sigma ref} \cdot \vec{\Omega}_{\sigma ref} / n_\sigma - \vec{R}_\sigma \cdot \vec{\Omega}_\sigma / n_\sigma)$ to explicitely show that friction dissipates canonical helicity. The second term on the right-hand side is the generalization of the "battery" $\int \phi\, \vec{B} \cdot d\vec{S}$ term in the evolution of magnetic helicity [14] and can be explicitly written as

$$\dot{K}_{\sigma rel}^{batt} = 2 \int_S (h_\sigma - h_{\sigma ref}) \vec{\Omega}_\sigma \cdot d\vec{S} \qquad (7)$$

to explicitely show that non-uniform enthalpy impose on the surface boundary of canonical vorticity flux tubes injects canonical helicity. The third term on the right-hand side is the generalization of the inductive $\int \vec{A}_+ \times \partial \vec{A}_- / \partial t \cdot d\vec{S}$ terms in the evolution of relative magnetic helicity to canonical helicity injection with time-dependent magnetic fields or mechanical torques $\vec{u}_\sigma \times \partial \vec{u}_\sigma / \partial t$. For example, in a simple neutral fluid flow, this term represents vortex generation due to an imposed torque in the linear flow. Canonical vorticity $\vec{\Omega}_\sigma$ plays an analogous role in two-fluid plasmas as the magnetic field $\vec{B}$ does in center-of-mass magnetohydrodynamics (MHD).

## 5. Two-fluid model of flux rope merging

The model for flux rope merging involves two reconnecting canonical flux ropes. Since canonical flux ropes (surfaces of constant $\Psi_\sigma \equiv \int \vec{\Omega}_\sigma \cdot d\vec{S}$) are effectively the linear superposition of magnetic flux ropes (surfaces of constant $\psi \equiv \int \vec{B} \cdot d\vec{S}$) with flow vorticity flux ropes (surfaces of constant $f_\sigma \equiv \int \vec{\omega}_\sigma \cdot d\vec{S}$), there are six possible merging combinations corresponding to the merging of two canonical flux ropes: a magnetic flux rope with another magnetic flux rope, a magnetic flux rope with an ion or electron flow vorticity flux rope, and merging between ion or electron flow vorticity flux ropes. In the reduced two-fluid regime where electron inertial effects are ignored ($m_e \rightarrow 0$), electron canonical flux ropes are geometrically indistinguishable from magnetic flux ropes, reducing the analysis to three possible merging scenarios (figure 2). During the merging of two canonical flux ropes, a common surface $S_{sep}$ separates both flux ropes on which $\vec{\Omega}_i \cdot d\vec{S} = \vec{\Omega}_e \cdot d\vec{S}$. Supposing that the overall system volume $V$ bounded by $S$ is isolated, the surface integrals in equations (6) and (7) are not performed over the overall surface boundary $S$ but only on the separation surface $S_{sep}$, because the canonical vorticity flux ropes do not penetrate the boundary $S$ but do penetrate the surface $S_{sep}$. Since the overall system is isolated, it is appropriate to take a species' enthalpy as the reference for the other species ($h_{\sigma ref} = h_\alpha$), and substituting into equation (7) shows that the total canonical helicity $\mathbb{K} = \mathbb{K}_{rel} = K_{irel} + K_{erel}$ is conserved and canonical helicity from one species can be transferred to the other species provided there is an enthalpy difference or a time-dependent induction on a finite common separation surface between the two species' flux ropes.

Figure 2 shows a geometric interpretation of how canonical helicity can be transferred between flux ropes. Consider two magnetic flux tubes without any flow that cross over each other (figure 2a). The end boundary conditions are the same as Refs. [1, 2, 21, 26], namely, they are such that each flux tube has no self-helicity and only a mutual helicity. The ends of the flux tubes can be closed (as in figure 1 of [1]) or one open and one closed as in figure 3.8 of Ref. [2] or any other combination. Since the actual number of helicities is not important but only the transfer of helicity is important, the ends of the flux tubes of figure 2 are left open for clarity to emphasize the creation of twists (self-helicity) under helicity conservation at the reconnection region just as in figure 2 of Ref. [26]. The total initial magnetic helicity of the system is taken to be one unit of helicity [1, 2, 21, 26], or $\mathcal{K}^{initial} = 1$. This is exactly equivalent to two ion canonical flux tubes that cross over each other (figure 2b), where $K_i^{initial} = \mathcal{H}^{initial} + \mathcal{X}^{initial} + \mathcal{K}^{initial} = 0 + 0 + 1$, without any flow so $\mathcal{H}^{initial} = 0$ and $\mathcal{X}^{initial} = 0$. Physical constants are dropped here for clarity. Upon merging or reconnection, one of three possible scenarios can occur. The first scenario (figure 2c) results in the familiar two magnetic flux ropes with a half-twist each, no crossing and no flow. In this case the canonical flux ropes are still just the magnetic flux ropes $\vec{\Omega}_i = \vec{B}$. Magnetic helicity is preserved with each flux rope having half a unit of magnetic helicity $\mathcal{K}_1^{final} = 1/2$ and $\mathcal{K}_2^{final} = 1/2$, and no flow $\mathcal{H}_1^{final} = 0$ and $\mathcal{H}_2^{final} = 0$, conserving the total final canonical helicity $K_i^{final} = 0 + 0 + 1/2 + 1/2 = K_i^{initial}$. The second scenario (figure 2d) results in a magnetic flux rope with half a twist and no flow, and a magnetic flux rope with no twist but a flow vorticity with half a twist. The total magnetic helicity is partially annihilated, $\mathcal{K}_1^{final} = 0$ (magnetic flux rope 1 does not twist) and $\mathcal{K}_2^{final} = 1/2$ (magnetic flux rope 2 twists a half turn). Magnetic helicity is partially converted to flow vorticity in flux rope 1, $\mathcal{H}_1^{final} = 1/2$, such that the total final canonical helicity is conserved, $K_i^{final} = 1/2 + 0 + 1/2 = K_i^{initial}$. The third scenario (figure 2e) involves total magnetic helicity annihilation, $\mathcal{K}_1^{final} = 0$ and $\mathcal{K}_2^{final} = 0$ (neither magnetic flux ropes twist) but flow vorticities develop in each flux rope, $\mathcal{H}_1^{final} = 1/2$ and $\mathcal{H}_2^{final} = 1/2$, while still preserving the total ion canonical helicity $K_i^{final} = 1/2 + 1/2 + 0 + 0 = K_i^{initial}$. The mechanism for helicity transfer is the enthalpy difference at the reconnection surface $S_{sep}$ as shown in figures 1a and 2b and described by equations (6) and (7). Which scenario occurs is revealed by taking the ratio of two terms in the expanded version of equation (7)

$$\frac{m_i \Delta h \int \vec{\omega}_i \cdot d\vec{S}}{q_i \Delta h \int \vec{B} \cdot d\vec{S}} \sim \frac{\rho_{Li}}{L} \tag{8}$$

assuming that enthalpy differences can be taken out of the integral, and velocities are of the order of the thermal and Alfvén velocities over the scale length $L$, and the thermal ion gyroradius $\rho_{Li} = m_i v_{ith}/q_i |\vec{B}|$. Equation (8) shows that for a given enthalpy imposed on the separation surface between two flux ropes, if the scale length is greater than the the ion scale length then helicity is channeled primarily into the magnetic component, while for scale lengths less than the ion scale length, helicity is channeled increasingly into the flow component. The ion gyroradius scale length normalized to a given scale length is also known as the size parameter [10] $S^* = \rho_{Li}/L$ or the ion skin depth $c/\omega_{p\sigma}$ where $c$ is the speed of light and $\omega_{p\sigma}$ is the plasma frequency.

Experiments have observed the dependence on the ion scale length of helicity channeling into the flow or the magnetic component during merging of flux ropes [10, 9, 8]. For example, the TS-4 experiment has observed bifurcation of compact torus formation from merging of counter-helicity spheromaks [10], with the bifurcation threshold depending on equation (8) [24]. In two-fluid flowing equilibria of compact plasmas [16], a field-reversed configuration (FRC) corresponds to a minimum energy flux rope with finite ion canonical helicity but no electron canonical helicity ($K_i \neq 0, K_e = \mathcal{K} = 0$), and a spheromak corresponds to a minimum energy flux rope with only finite magnetic helicity. Any canonical helicity preferentially channeled into the vortex component ($m_i \cancel{f} \gg q_i \psi$) results in an FRC and any canonical helicity preferentially channeled into the magnetic component with negligible vortex component ($m_i \cancel{f} \ll q_i \psi$) results in a spheromak. Figure 3 shows the helicity accounting for compact torus formation in both cases and represents a geometric interpretation of the TS-4 experimental results. The two initial spheromaks have opposite helicities, $\mathcal{K}_L = \psi_L^2$ for the left spheromak and $\mathcal{K}_R = -\psi_R^2$ for the right spheromak, with the ratio of helicities $r = |\mathcal{K}_R|/\mathcal{K}_L$ controllable by the operator. Therefore the total initial magnetic helicity in the system is $\mathcal{K}^{init} = \mathcal{K}_L(1-r) \neq 0$. If $r$ is less than a threshold, the final compact torus formed from merging of the two initial spheromaks is an FRC with $\mathcal{K}^{final} \sim 0$. Magnetic helicity appears to have been annihilated. If $r$ is greater than a threshold, the final compact torus formed is a spheromak with magnetic helicity $\mathcal{K}^{final} \simeq \mathcal{K}^{init}$. The actual details involves MHD activity with mode cascades that convert flux at high $L/\rho_{Li}$ to compensate for toroidal flux annihilation, or shear flow generation that prevents MHD activity and flux conversion at low $L/\rho_{Li}$. From a canonical helicity point-of-view, the interpretation is simple (figure 3). The initial ion helicity is $K_{iL} = \mathcal{K}_L$ and $K_{iR} = \mathcal{K}_R$, therefore the total initial canonical helicity of the system is $\mathbb{K}^{init} = 2\mathcal{K}_L(1-r)$. After merging, when $r$ is below the threshold, the total magnetic helicity $\mathcal{K} = \mathcal{K}_L(1-r)$ is converted to ion canonical helicity so $K_i^{final} = 2\mathcal{K}_L(1-r)$. So magnetic helicity has not been annihilated but converted in such a way into flows without magnetic twist (FRC) as to still preserve the total canonical helicity $\mathbb{K}^{final} = \mathbb{K}^{init}$. When $r$ is above the threshold, magnetic helicity is conserved as usual, as is total canonical helicity. The two-fluid canonical helicity transport model explains why the choice of one path or another depends on the scale lengths over which enthalpy is imposed on the merging system, compared to the ion scale lengths, and controls the channelling of helicity into the vorticity flux or the magnetic flux component of the canonical flux rope.

## 6. Relationship between MHD magnetic helicity and two-fluid helicities

The transport equations for canonical helicity can be derived equivalently from the MHD point-of-view or from the two-fluid point-of-view. From the MHD point-of-view, it may be more familiar (but more difficult) to begin with the transport equation for relative magnetic helicity, written as

$$q_\sigma^2 \frac{d\mathcal{K}_{rel}}{dt} = -2q_\sigma^2 \int_{V_\sigma} \vec{E} \cdot \vec{B} \, dV + 2q_\sigma^2 \int \phi \vec{B} \cdot d\vec{S} - q_\sigma^2 \int \vec{A} \times \frac{\partial \vec{A}}{\partial t} \cdot d\vec{S} \tag{9}$$

where the Leibniz convection term has been ignored here for convenience (assuming fixed boundaries). Instead of replacing the electric field $\vec{E}$ with a form of Ohm's law, the two-fluid equations of motion suitably re-arranged are inserted into equation (9) to provide the cross-helicity terms

$$\begin{aligned}
m_\sigma q_\sigma \frac{d\mathcal{X}_\sigma}{dt} = &\, 2q_\sigma^2 \int \vec{E} \cdot \vec{B} \, dV - 2q_\sigma \int \vec{B} \cdot \frac{\vec{R}_\sigma}{n_\sigma} dV - 2q_\sigma \int \vec{B} \cdot \frac{\nabla \mathcal{P}_\sigma}{n_\sigma} dV \\
&- m_\sigma q_\sigma \int u_\sigma^2 \vec{B} \cdot d\vec{S} - 2m_\sigma q_\sigma \int \vec{\omega}_\sigma \cdot (\vec{u}_\sigma \times \vec{B}) \, dV \\
&- 2m_\sigma q_\sigma \int \vec{u}_\sigma \cdot \nabla \times \vec{E} \, dV
\end{aligned} \tag{10}$$

and the kinetic helicity terms

$$\begin{aligned}
m_\sigma^2 \frac{d\mathcal{H}_\sigma}{dt} = &\, -m_\sigma q_\sigma \int \vec{u}_\sigma \times \frac{\partial \vec{A}}{\partial t} \cdot d\vec{S} + 2m_\sigma q_\sigma \int \vec{\omega}_\sigma \cdot (\vec{u}_\sigma \times \vec{B}) dV \\
&- 2m_\sigma \int \vec{\omega}_\sigma \cdot \frac{\vec{R}_\sigma}{n_\sigma} dV - 2m_\sigma q_\sigma \int \vec{u}_\sigma \cdot \frac{\partial \vec{B}}{\partial t} dV \\
&- m_\sigma^2 \int \vec{u}_\sigma \times \frac{\partial \vec{u}_\sigma}{\partial t} \cdot d\vec{S} - m_\sigma q_\sigma \int \vec{A} \times \frac{\partial \vec{u}_\sigma}{\partial t} \cdot d\vec{S} \\
&- 2m_\sigma \int h_\sigma \vec{\omega}_\sigma \cdot d\vec{S}
\end{aligned} \tag{11}$$

which all add up to be equivalent to equation (2). Note that it is a little simpler to begin with the two-fluid formulation and transform to the center-of-mass (MHD) frame, where MHD kinetic helicity is related to the species' kinetic helicity by $\mathcal{H} = (n_i m_i \mathcal{H}_i + n_e m_e \mathcal{H}_e)/nm$ with $nm = n_i m_i + n_e m_e$ and the total derivatives are re-defined using the convection for each species $d/dt = \partial/\partial t + \vec{u}_\sigma \cdot \nabla$. Terms involving $\nabla \times \vec{J}$ are included at the appropriate steps. These terms are sources of electromagnetic waves due to rotation of current density (vorticity of charged fluid elements) in the inhomogeneous electromagnetic wave equation, and are neglected in the MHD approximation because only slow phenomena are considered. These terms, most easily derived from the two-fluid point-of-view first, show that kinetic helicity can be changed by electromagnetic waves (e.g. RF current drive [27]).

These expressions provide a geometric interpretation of the mechanisms for coupling and transfer between magnetic, cross and kinetic helicities. The parallel electric field in equation (9) is a sink of magnetic helicity, but becomes a source of cross-helicity in equation (10). In turn, the last two terms on the right-hand side of equation (10) are sinks of cross-helicity but sources of kinetic helicity in equation (11). Kinetic helicity decays with friction parallel to the flow vorticity in the volume and cross-helicity decays with friction parallel to the magnetic field in the volume. Taking ratios of the appropriate terms then provides the conditions under which magnetic helicity is converted into cross-helicity, then decays or is in turn converted into kinetic helicity (flows). This kinetic helicity can then decay with friction. The converse could be interpreted as a two-fluid geometric picture of the dynamo effect, where kinetic helicity can couple and transfer to cross-helicities and magnetic helicity by suitable enthalpy at the boundaries.

## 7. Summary

A two-fluid model for the merging of flux-ropes based on the transport of canonical helicity has been developed. The mechanisms for helicity transfer can be via non-uniform enthalpy across a common separation surface between two canonical flux ropes, time-dependent magnetic fields or mechanical torques in a flow field. Because canonical flux ropes are the superposition of magnetic flux ropes and flow vorticity flux ropes of each species, the model can work with any of the possible merging combinations: magnetic with magnetic, magnetic with flows and flows with flows. The model retrieves all previous helicity conservation solutions in the appropriate limits: magnetic helicity conservation when mass or friction is negligible, neutral flows helicity conservation when Lorentz forces are neglected, species self-helicity conservation when separation surfaces or enthalpy non-uniformities are neglected. The model also provides conditions for the direction of helicity transfer. In particular when the scale lengths of enthalpy non-uniformities are smaller than the ion skin depth, then magnetic helicity is converted to helical ion flows, and when larger, magnetic helicity is preserved. This model provides an intuitive explanation for experimental observations of compact torus bifurcation during merging-reconnection. Geometric interpretations for canonical helicity transfer has also been provided. In summary, the two-fluid helicity transport model for flux-rope merging provides a general but still intuitive interpretation for the dynamics of flux ropes with or without strong plasma flows.

**Acknowledgements**

This work was supported in part by the DOE Grant number DE-SC0010340. The author would like to thank the referees for pointing out Refs. [12, 19, 25] and help on improving the manuscript.

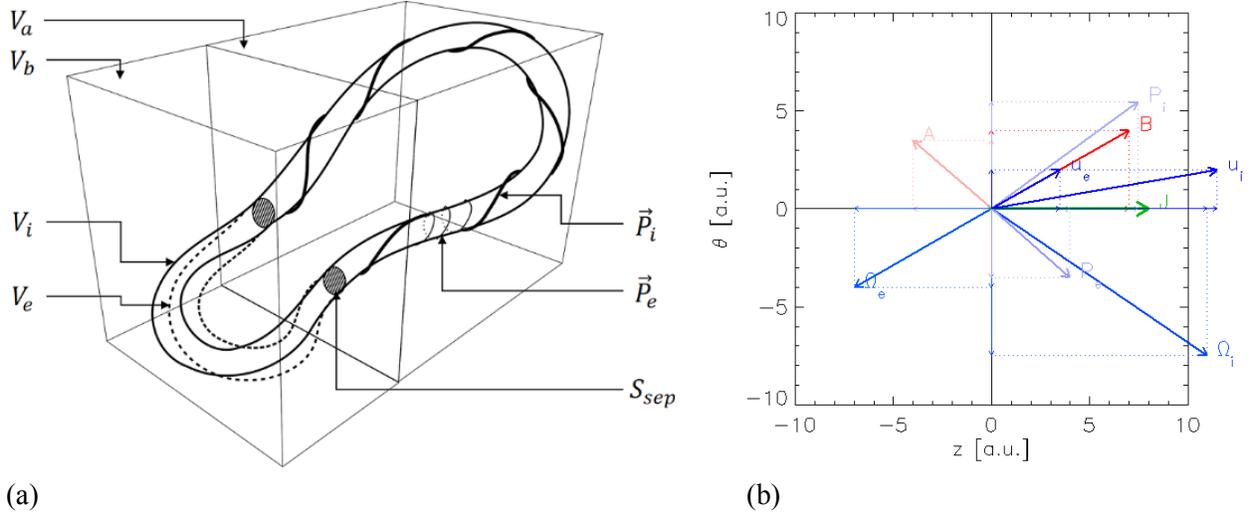

(a) (b)

**Figure 1.** Geometric interpretation of canonical helicity. (a) Schematic of ion and electron canonical flux ropes $V_i, V_e$ defined by the canonical momentum $\vec{P}_i, \vec{P}_e$. In volume $V_a$ both flux ropes are not distinguishable, for example at to large scales, while in volume $V_b$, both flux ropes are distinguishable. The separation surface $S_{sep}$ delimits the common surface of each flux tubes through which a non-uniform enthalpy can transfer canonical helicity (equation (8)). (b) Unwrapping an arbitrary surface of a cylindrical magnetic flux rope with axial current density $\vec{J} = J\,\hat{z}$ resulting in helical magnetic fields $\vec{B}$. Supposing $m_e \to 0$, the electron flow velocity $\vec{u}_e$ must be parallel (or anti-parallel) to $\vec{B} = \nabla \times \vec{A}$ and the ion flow must be such that the azimuthal ion flow is equal to the electron azimuthal flow (no net azimuthal current). This results in helical electron and ion canonical momenta $\vec{P}_e, \vec{P}_i$, and helical canonical vorticity $\vec{\Omega}_e, \vec{\Omega}_i$ with finite canonical helicity.

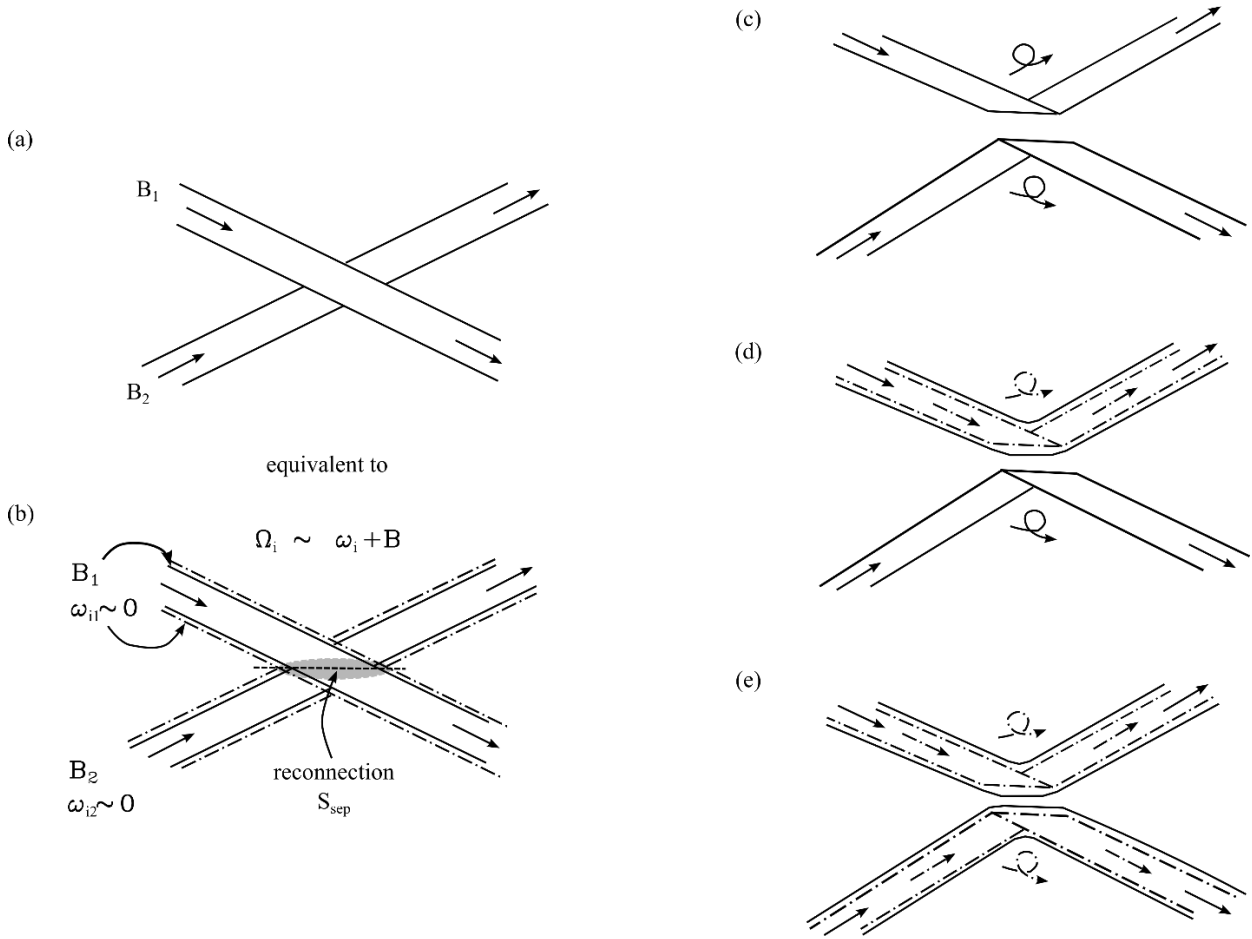

**Figure 2.** Merging of canonical flux ropes (reduced two-fluid regime). (a) Magnetic flux ropes with one crossing, giving a total initial magnetic helicity of 1 unit [2, 21, 26], ie. $\{K_i; \mathcal{H}_i; \mathcal{X}_i; \mathcal{K}\} = \{1; 0; 0; 1\}$. This is equivalent to (b) two crossing canonical flux ropes with no flow vorticity $\vec{\omega}_i \sim 0$. Merging reconnection occurs at a separation surface $S_{sep}$ with a non-uniform enthalpy into equation (7) that can result in one of three scenarios: (c) two canonical flux ropes with no flow but a half-twist magnetic flux, i.e. two magnetic flux ropes, conserving total canonical helicity and magnetic helicity $\{K_i; \mathcal{H}_i; \mathcal{X}_i; \mathcal{K}\} = \{1; 0; 0; 1/2 + 1/2\}$; (d) two canonical flux ropes, one with a half-twist flow vorticity but no magnetic twist and the other with a half-twist magnetic flux but no flow twist, after converting part of the magnetic helicity in the initial crossing into flow twist, conserving total canonical helicity $\{K_i; \mathcal{H}_i; \mathcal{X}_i; \mathcal{K}\} = \{1; 1/2; 0; 1/2\}$; (e) two canonical flux ropes, each with a half-twist flow vorticity and no magnetic twist, still conserving total canonical helicity $\{K_i; \mathcal{H}_i; \mathcal{X}_i; \mathcal{K}\} = \{1; 1/2 + 1/2; 0; 0\}$. Which scenario occurs is mediated by the relative strength of the terms in equations (8)-(11).

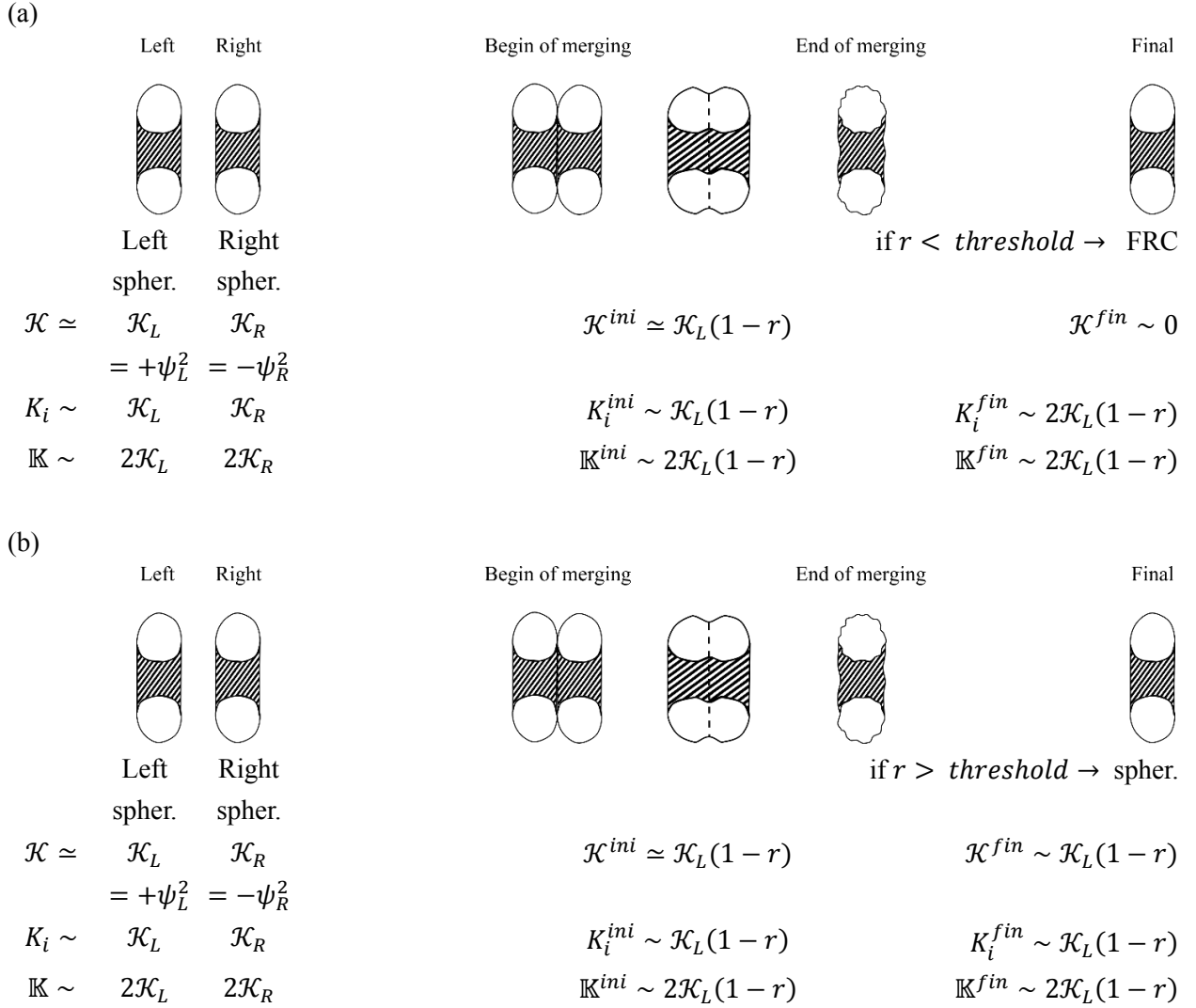

**Figure 3.** Helicity accounting during bifurcation in compact torus formation [10, 24]. Two counter-helicity spheromaks are formed, with the right spheromak helicity experimentally varied by the operator with a ratio $r = |\mathcal{K}_R|/\mathcal{K}_L$. (a) If the helicity ratio $r$ is below a threshold, the final compact torus configuration is an FRC with negligible magnetic helicity $\mathcal{K}^{fin} \ll \mathcal{K}^{ini}$. From a canonical helicity point-of-view, the initial magnetic helicity has been converted to ion canonical helicity $K_i^{fin} > K_i^{ini}$ while preserving the total canonical helicity of the system $\mathbb{K}^{fin} \simeq \mathbb{K}^{ini}$. (b) If the helicity ratio $r$ is above a threshold, the final configuration is a spheromak without magnetic helicity conversion. The threshold has been experimentally measured to be proportional to the ion size parameter $L/\rho_{Li}$ and explained by the two-fluid canonical helicity merging model equation (8).